\documentclass[12pt,english,floatfix,superscriptaddress,aps,preprint,showpacs]{revtex4}
\usepackage{amsmath}
\usepackage{amssymb}
\usepackage{amsbsy}
\usepackage{amsfonts}
\usepackage{amsopn}
\usepackage{amstext}
\usepackage{graphicx}
\usepackage{esint}
\usepackage{amssymb}
\usepackage{amsfonts}
\usepackage{amsmath}
\usepackage{graphicx}
\usepackage[english]{babel}
\usepackage{color}
\usepackage[dvips]{epsfig}
\usepackage[dvips]{graphicx}
\usepackage{float}
\usepackage{units}
\usepackage{textcomp}
\usepackage{babel}
 \usepackage{hyperref}
\begin{document}

\title{Einstein-Hilbert graviton modes modified by the\ Lorentz-violating
bumblebee Field}
\author{R. V. Maluf}
\email{r.v.maluf@fisica.ufc.br}
\affiliation{Departamento de F\'{\i}sica, Universidade Federal do Cear\'{a} (UFC), Campus
Universit\'{a}rio do Pici, Caixa Postal 6030, 60455-760 - Fortaleza, Cear\'{a}, Brazil}
\author{C.~A.~S.~Almeida}
\email{carlos@fisica.ufc.br}
\affiliation{Departamento de F\'{\i}sica, Universidade Federal do Cear\'{a} (UFC), Campus
Universit\'{a}rio do Pici, Caixa Postal 6030, 60455-760 - Fortaleza, Cear\'{a}, Brazil}
\author{R.~Casana}
\email{rodolfo.casana@gmail.com}
\affiliation{Departamento de F\'{\i}sica, Universidade Federal do Maranh\~{a}o (UFMA),
Campus Universit\'{a}rio do Bacanga, 65085-580, S\~{a}o Luis, Maranh\~{a}o, Brazil}
\author{M.~M.~Ferreira~Jr}
\email{manojr.ufma@gmail.com}
\affiliation{Departamento de F\'{\i}sica, Universidade Federal do Maranh\~{a}o (UFMA),
Campus Universit\'{a}rio do Bacanga, 65085-580, S\~{a}o Luis, Maranh\~{a}o, Brazil}

\begin{abstract}
In this work, we investigate the consequences of the spontaneous breaking of
Lorentz symmetry, triggered by the bumblebee vector field, on the
usual Einstein-Hilbert theory. Specifically, we consider the
Einstein-Hilbert action modified by the bumblebee dynamic field, and
evaluate the graviton propagator using an extended basis of Barnes-Rivers
tensor projectors, involving the Lorentz-violating vector. Once the
propagator is carried out, we proceed with discussing the consistency of the
model, writing the dispersion relations, and analyzing causality and unitarity.
We verify that this model possesses two dispersion relations: one
provides causal and unitary propagating modes, while the second yields a
causal but nonunitary mode which spoils the physical consistency of the
model.
\end{abstract}

\pacs{11.30.Cp, 04.25.Nx, 12.60.-i}
\maketitle

\section{Introduction}

Theories with Lorentz-symmetry breaking have been under intensive
investigation since the proposal of the standard model extension (SME) \cite%
{Colladay1,Colladay2,Coleman,Samuel1,Samuel2,Samuel3,Samuel4,Potting1,Potting2,Potting3}
as a broader version of the usual standard model incorporating tensor terms
generated by spontaneous Lorentz violation. The Lorentz-violating (LV)
terms, generated as vacuum expectation values of tensors defined in a high
energy scale, are coupled to the physical fields yielding coordinate
invariance and violation of Lorentz symmetry in the particle frames \cite%
{Lehnert}. This theoretical framework has inspired a large number of
investigations in the last several years, encompassing fermion systems \cite%
{fermion1,fermion2,fermion3,fermion4,fermion5,fermion6,fermion7,fermion8,fermion9}%
, CPT-probing experiments \cite{CPT1,CPT2,CPT3,CPT4,CPT5, CPT6,CPT7}, the
electromagnetic CPT- and Lorentz-odd term \cite{Adam1}-\cite{Gaete}, the
CPT-even and Lorentz-odd gauge sector and its interactions with fermions
\cite{KM1,KM2,KM3,Risse1,Risse2,Risse3,Risse4,Risse5,Schreck1,Barone1}, \cite%
{Baeta2,Baeta3, Baeta4,Barone2,Silva1,Silva2}. Recent investigations
involving higher dimensional operators \cite{Kostelec1,Kostelec2, Kostelec3}%
, its possible connections with LV theories \cite%
{Myers1,Reyes,Marat1,Marat2, Marat3,Marat4}, and nonminimal couplings \cite%
{NModd1,NModd2,NM1,NM2,NM3,NM4}, have also been reported.

The interest in an extension of the SME embracing gravity comes from the
fact that Lorentz violation may be a key ingredient of a quantum theory for
gravitation. Indeed, Lorentz-violating effects might be significant
in regions or situations were the curvature or torsion are large, as in the
vicinity of black holes. Furthermore, these effects may also play relevant
role in cosmological scenarios described by dark energy or dark matter, or
the ones where anisotropy factors can be inserted in the
Friedman-Robertson-Walker solutions. Lorentz violation in the gravitational
sector may be theoretically investigated in connection with tests sensitive
to the inverse square law, the deflection of light, geodesic precession,
between others. A consistent formalism to include LV terms in gravity
requires a framework compatible with non-null vacuum expectation values that
break local Lorentz symmetry but keeps the general coordinate invariance.
The Riemann-Cartan geometry, endowed with dynamic curvature and torsion, was
used for such a purpose in Ref. \cite{KosteleckyG1}, where the LV coupling
terms were constructed using vierbein and spin connections. In Ref. \cite%
{Bluhm2005}, the connection between Nambu-Goldstone modes and the
spontaneous violation of local Lorentz and diffeomorphism symmetries were
investigated in the Riemann-Cartan spacetime using the vierbein and spin
connection formalism previously developed. In Ref. \cite{Bailey2006},
signals for Lorentz violation in post-Newtonian gravity were scrutinized in
the case of a Riemann spacetime (null torsion) by considering the linearized
Einstein equations modified by 20 independent dynamical LV coefficients
generated by spontaneous symmetry breaking. New developments were performed
in Refs. \cite{Potting2009,Tasson2011}. Also, alternative approaches for
Lorentz violation in curved space, focused on a more geometric point of
view, have been discussed in Refs. \cite{Euclides1,Euclides2,Euclides3}.
Investigations about Lorentz-violating linearized gravitation \cite{Petrov}
and high order gravity models modified by Lorentz-violating terms were also
reported \cite{Pereira2011, Boldo}.

In accordance with these studies the extension of the gravitational sector
including Lorentz-violating terms is given by the action
\begin{equation}
S=S_{EH}+S_{LV}+S_{matter,}
\end{equation}
\ where $S_{EH}$\ represents the usual Einstein-Hilbert action,
\begin{equation}
S_{EH}=\int d^{4}x\sqrt{-g}\frac{2}{\kappa^{2}}\left( R-2\Lambda\right) ,
\label{SEH}
\end{equation}
where $R$ is the curvature scalar and $\Lambda$ is the cosmological constant.
Moreover, the action\ $S_{LV}$\ accounts for Lorentz-violating leading
terms, written as
\begin{equation}
S_{LV}=\!\!\int\!\!d^{4}x\sqrt{-g}\frac{2}{\kappa^{2}}\left( uR+s^{\mu\nu
}R_{\mu\nu}+t^{\mu\nu\alpha\beta}R_{\mu\nu\alpha\beta}\right) ,  \label{SLV1}
\end{equation}
with $u$, $s^{\mu\nu}$ and $t^{\mu\nu\alpha\beta}$ being tensors which
enclose the Lorentz-violating coefficients, and $\kappa^{2}=32\pi G$\ being
the gravitational coupling. The dimensionless tensors $s^{\mu\nu}$, $t^{\mu
\nu\alpha\beta}$ possess the same symmetries of the Ricci and Riemann
tensors, respectively, and are to be considered as traceless, $%
s^{\mu}{}_{\mu}=$\ $t^{\mu\nu}{}_{\mu\nu}=0$, once their traces can be
absorbed in the scalar $u.$\ Moreover, the components $t^{\mu\nu\beta}{}_{%
\nu}$ can be also taken as null once they can be absorbed in the tensor $%
s^{\mu\beta}$. So, the tensors $s^{\mu\nu}$, $t^{\mu\nu\alpha\beta}$ have nine and ten independent components, respectively.

The bumblebee model is a simple example of gravity model where a vector
field $B^{\mu}$\ acquires a nonzero vacuum expectation value inducing
Lorentz and diffeomorphism violations. This model was first considered in
the context of string theories \cite{Samuel1}, with the spontaneous Lorentz-symmetry breaking being triggered by the potential $V(B^{\mu})=\lambda\left(
B^{\mu}B_{\mu}\mp b^{2}\right) ^{2}/2.$ In accordance with the literature
\cite{KosteleckyG1,Bluhm2005}, the vector bumblebee model can be represented
as stated in action (\ref{SLV1}), whenever $t^{\mu\nu\alpha\beta}=0$ and%
\begin{equation}
u=\frac{1}{4}\xi B^{\alpha}B_{\alpha},\text{ \ }s^{\mu\nu}=\xi\left( B^{\mu
}B^{\nu}-\frac{1}{4}g^{\mu\nu}B^{\alpha}B_{\alpha}\right) ,  \label{LVT1}
\end{equation}
with $s^{\mu\nu}$ being traceless. With such definitions, the action
responsible for the dynamics of the bumblebee field $B_{\mu}$ is written as
\begin{equation}
S_{B}=\int d^{4}x\sqrt{-g}\left[ -\frac{1}{4}B^{\mu\nu}B_{\mu\nu}+ \frac{%
2\xi }{\kappa^{2}}B^{\mu}B^{\nu}R_{\mu\nu} - V(B^{\mu}B_{\mu}\mp b^{2})%
\right] ,  \label{SLV2}
\end{equation}
where we have included the\ corresponding field strength%
\begin{equation}
B_{\mu\nu}=\partial_{\mu}B_{\nu}-\partial_{\nu}B_{\mu},
\end{equation}
and the quadratic potential,
\begin{equation}
V=\frac{\lambda}{2}\left( B^{\mu}B_{\mu}\mp b^{2}\right) ^{2}.
\end{equation}
that triggers the spontaneous breakdown of diffeomorphism symmetry. Here, $b^{2}$\ is a positive constant that stands for the nonzero vacuum
expectation value of this field. \ All quantities are expressed in natural
units $(\hslash=c=\epsilon_{0}=1)$, including the gravitational constant, $G=6.707\times10^{-57}\mbox{(eV)}^{-2}$, so that the mass dimension of constants and
fields are $\left[ \kappa^{2}\right] =-2,\left[ B^{\mu}\right] =1,\left[
B^{\mu\nu}\right] =2,$ $\left[ \lambda\right] =0,\left[ \xi\right] =-2.$ The
constant $\xi$\ is the one that establishes the nonminimal coupling between
the bumblebee field and the curvature tensor, keeping $u$\ and $s^{\mu\nu}$\
dimensionless. Moreover, tensors are symmetrized with unit weight, i.e., $A_{(\mu\nu)}=\frac{1}{2}(A_{\mu\nu}+A_{\nu\mu})$.

In Ref. \cite{Bailey2006}, the effects of the linearized version of the
bumblebee model on the Einstein-Hilbert gravity were analyzed.\ This model
was also addressed in Refs. \cite{Potting2009,Tasson2011,BluhmFung}. \ Some
additional implications of this model on the Newtonian gravitational
potential have been recently addressed in Ref. \cite{MalufGravity1}, where
the weak-field formalism of gravity was used to calculate the bumblebee
corrections induced on the gravitational potential. It was then shown that
the coupling of this field with the curvature tensor, as stated in action (%
\ref{SLV2}), implies an anisotropic potential correction proportional to $%
b_{i}b_{j}\hat{x}^{i}\hat{x}^{j}$, which breaks the spatial isotropy of the
ordinary gravitational potential. Besides, an
additional correction similar to the well known electric Darwin term, $%
\nabla^{2}\frac {1}{r}\sim\delta^{(3)}(\vec{x})$ was also reported, giving rise to a very weak
and short-ranged contribution to the gravitational interaction.

In spite of the fact that the results of Ref. \cite{MalufGravity1} are
consistent with the literature \cite{Bailey2006,Tasson2011}, they are based on
a preliminary form of the Einstein-Hilbert graviton propagator modified by
the linearized bumblebee field, evaluated as a perturbative insertion on the
usual case. This approach, however, does not provide an exact result, being
not suitable to analyze the vacuum structure of this model and the
properties of the physical excitations around it. It is known that the
spontaneous Lorentz breaking is always accompanied by diffeomorphism
violation \cite{Bluhm2005,Bailey2006,Potting2009}, so the graviton spectrum
may undergo nontrivial modifications, as the generation of massive modes,
the appearance of nonphysical modes concerning causality (tachyons) and
unitarity aspects (ghosts).

In this work, we intend to investigate the graviton spectrum in the context
of the linearized Einstein-Hilbert gravity (without torsion)
endowed with the spontaneous violation of Lorentz symmetry\ induced by the
bumblebee field, as studied in Ref. \cite{MalufGravity1}.\ In this sense, we
exactly carry out the graviton propagator (in tree-level approximation)
applying a general method based on the Barnes-Rivers spin operators \cite%
{Barnes,Rivers,Sezgin} and recently extended\ in Ref. \cite{Boldo} for the
case of gravity theories with Lorentz-breaking terms.\ Once the graviton
Feynman propagator is evaluated, one also analyzes the consistency
(stability, causality, unitarity)\ of this theory starting from the
dispersion relations stemming from the poles of the propagator. In the
present work we use the spacetime signature $(+\ -\ -\ -)$ and adopt the
following definition for the Ricci tensor: $R_{\mu \nu }=\partial _{\sigma
}\Gamma _{\mu \nu }^{\sigma }-\partial _{\nu }\Gamma _{\mu \sigma }^{\sigma
}+\Gamma _{\sigma \lambda }^{\lambda }\Gamma _{\mu \nu }^{\sigma }-\Gamma
_{\sigma \nu }^{\lambda }\Gamma _{\mu \lambda }^{\sigma },$ where $\Gamma
_{\mu \nu }^{\lambda }=\frac{1}{2}g^{\lambda \sigma }\left( \partial _{\mu
}g_{\nu \sigma }+\partial _{\nu }g_{\mu \sigma }-\partial _{\sigma }g_{\mu
\nu }\right) .$

The structure of the paper is as follows. In Sec. \ref{sec:theoretical-model}
we present the linearized bilinear gravity action for which we evaluate the
Feynman propagator, using an extended basis of the Barnes-Rivers projectors. In
Sec. III, we present the dispersion relations coming from the poles of the
propagator, and discuss the stability and causality issues. The unitarity
analysis is investigated in Sec. IV, while our concluding comments are
presented in Sec. \ref{sec:Conclusions}.

\section{Theoretical model and the graviton propagator\label%
{sec:theoretical-model}}

In order to determine the influence of the gravity-bumblebee coupling on the
graviton dynamics,\ we consider the actions (\ref{SLV1}) and (\ref{SLV2}),
following the route described in Refs.\cite%
{Bailey2006,Tasson2011,MalufGravity1}. For assessing the linearized version,
we split the dynamic fields into the vacuum expectation values and the
nearby quantum fluctuations:
\begin{align}
g_{\mu\nu} & =\eta_{\mu\nu}+\kappa h_{\mu\nu},  \notag \\
B_{\mu} & =b_{\mu}+\tilde{B}_{\mu},  \label{eq:Expansion1} \\
B^{\mu} & =b^{\mu}+\tilde{B}^{\mu}-\kappa b_{\nu}h^{\mu\nu},  \notag
\end{align}
where $h_{\mu\nu}$ and $\tilde{B}_{\mu}$ represent small perturbations
around the Minkowski background and a constant vacuum value $b_{\mu}$,
respectively. The quantity $b^{\mu}=(b_{0},\mathbf{b})$ represents the fixed
background responsible for the violation of Lorentz and $CPT$ symmetries in
the local frame of particles \cite{KosteleckyG1}.

Following the procedure outlined in Ref. \cite{Bailey2006}, the solution for
the linearized bumblebee equation of motion can be written in the momentum
space as
\begin{equation}
\tilde{B}^{\mu }=\frac{\kappa p^{\mu }b_{\alpha }b_{\beta }h^{\alpha \beta }%
}{2\left( b\cdot p\right) }+\frac{2\sigma b_{\alpha }R^{\alpha \mu }}{p^{2}}-%
\frac{2\sigma p^{\mu }b_{\alpha }b_{\beta }R^{\alpha \beta }}{p^{2}\left(
b\cdot p\right) }+\frac{\sigma p^{\mu }R}{4\lambda \left( b\cdot p\right) }-%
\frac{\sigma b^{\mu }R}{p^{2}}+\frac{\sigma p^{\mu }b^{2}R}{p^{2}\left(
b\cdot p\right) },  \label{Btilde}
\end{equation}%
with $p^{\mu }=(p_{0},\mathbf{p})$, $\sigma =(2\xi /\kappa ^{2})$, while $%
R_{\mu \nu }$ and $R$ are taken in linearized form. So, we can insert this
solution into the Lagrangian term representing the bumblebee interaction, $%
\mathcal{L}_{\text{LV}}=\sigma \sqrt{-g}B^{\mu }B^{\nu }R_{\mu \nu },$ to
determine the modifications implied by the background $b_{\mu }$ on the
kinetic sector of the graviton field $h_{\mu \nu }$, yielding the following
effective Lagrangian (already evaluated in Ref. \cite{MalufGravity1}):
\begin{eqnarray}
\mathcal{L}_{\text{LV}} &=&\xi \left[ p^{2}b_{\mu }b_{\nu }\left( h^{\mu \nu
}h+h^{\mu \alpha }h^{\nu }{}_{\alpha }\right) -\frac{1}{2}\left( b\cdot
p\right) ^{2}\left( h^{\mu \nu }h_{\mu \nu }-h^{2}\right) -\left( b_{\mu
}b_{\nu }p_{\alpha }p_{\beta }+b_{(\mu }p_{\nu )}b_{(\alpha }p_{\beta
)}\right) h^{\mu \nu }h^{\alpha \beta }\right]  \notag \\
&+&\frac{4\xi ^{2}}{\kappa ^{2}}\left[ \left( -2p^{2}b_{\mu }b_{\nu
}-2b^{2}p_{\mu }p_{\nu }+4\left( b\cdot p\right) b_{(\mu }p_{\nu )}-\frac{%
p^{2}p_{\mu }p_{\nu }}{4\lambda }\right) h^{\mu \nu }h\right.  \notag \\
&+&\left( 2b_{\mu }b_{\nu }p_{\alpha }p_{\beta }-b_{(\mu }p_{\nu
)}b_{(\alpha }p_{\beta )}+\frac{b^{2}p_{\mu }p_{\nu }p_{\alpha }p_{\beta }}{%
p^{2}}-\frac{2\left( b\cdot p\right) p_{\mu }p_{\nu }b_{(\alpha }p_{\beta )}%
}{p^{2}}+\frac{p_{\mu }p_{\nu }p_{\alpha }p_{\beta }}{4\lambda }\right)
h^{\mu \nu }h^{\alpha \beta }  \notag \\
&+&\left. \left( b^{2}p^{2}-\left( b\cdot p\right) ^{2}+\frac{p^{4}}{%
4\lambda }\right) h^{2}+\left( p^{2}b_{\mu }b_{\nu }-2(b\cdot p)b_{(\mu
}p_{\nu )}+\frac{\left( b\cdot p\right) ^{2}p_{\mu }p_{\nu }}{p^{2}}\right)
h^{\mu \lambda }h^{\nu }{}_{\lambda }\right] ,  \label{eq:LagrangianLV}
\end{eqnarray}%
where $h=h^{\alpha }{}_{\alpha }$. As expected, it is possible to verify
that the Lagrangian (\ref{eq:LagrangianLV}) is not invariant under the gauge
transformations $h_{\mu \nu }\rightarrow h_{\mu \nu }+ip_{\mu }\zeta _{\nu
}+ip_{\nu }\zeta _{\mu },$ for any arbitrary $\zeta _{\mu }$. Furthermore,
it is worth noting that there are second-order corrections $\mathcal{O(\xi }%
^{2})$ which introduce higher derivative terms, and are background
independent.

Another observation concerns the existing connection between the terms
involving the bumblebee field in the squared linearized Lagrangian (\ref%
{eq:LagrangianLV}) and the ones stemming from higher order Lagrangian terms,
as $\mathcal{L}=\beta R_{\mu\nu}R^{\mu\nu}+\gamma R^{2},$\ as depicted in
Refs. \cite{Pereira2011,Boldo,Accioly1,Accioly2,Accioly3}. We can show that
the linearized terms associated with $R^{2},$\ $R_{\mu\nu}R^{\mu\nu},$\
namely, $h\square^{2}h,$\ $h\square\partial_{\mu}\partial_{\nu}h^{\mu\nu },$%
\ $h^{\alpha\beta}\partial_{\alpha}\partial_{\beta}\partial_{\mu}\partial_{%
\nu}h^{\mu\nu},$\ $h^{\mu\nu}\square^{2}h^{\mu\nu}, $\ can be also found in
Lagrangian (\ref{eq:LagrangianLV}). This shows that the bumblebee field also
plays the role of inducing high order gravity terms on the Einstein-Hilbert
action.

Following the purpose of analyzing the effects of the bumblebee field on the
Einstein-Hilbert action, we should add the Lorentz-violating terms of Eq. (%
\ref{eq:LagrangianLV}) to the bilinear terms of the linearized
Einstein-Hilbert Lagrangian,
\begin{equation}
\mathcal{L}_{\text{EH}}=p_{\mu}p_{\alpha}h^{\mu\nu}h_{\ \nu}^{\alpha}-p_{\mu
}p_{\nu}h^{\mu\nu}h+\frac{1}{2}p^{2}h_{\ \mu}^{\mu}h-\frac{1}{2}p^{2}h^{\mu
\nu}h_{\mu\nu}.
\end{equation}
without introducing a gauge fixing-term. Our interest is the kinetic
Lagrangian,
\begin{equation}
\mathcal{L}_{\text{kin}}=\mathcal{L}_{EH}+\mathcal{L}_{\text{LV}}.
\label{L_E}
\end{equation}
To find the corresponding Feynman propagator for $\mathcal{L}_{\text{kin}},$
we first rewrite the resulting Lagrangian $\mathcal{L}_{\text{kin}}$ into
the bilinear form
\begin{equation}
\mathcal{L}_{\text{kin}}=-\frac{1}{2}h^{\mu\nu}\hat{\mathcal{O}}_{\mu
\nu,\alpha\beta}h^{\alpha\beta},  \label{eq:Operator}
\end{equation}
where the operator $\hat{\mathcal{O}}_{\mu\nu,\alpha\beta}$ is symmetric in
the indices $(\mu\nu)$, $(\alpha\beta),$ and under the interchange of the
pairs $(\mu\nu)$ and $(\alpha\beta)$. Following the notations and
conventions of Ref. \cite{MalufGravity1}, the graviton propagator is defined
as
\begin{equation}
\left\langle 0\left\vert T\left[ h_{\mu\nu}(x)h_{\alpha\beta}(y)\right]
\right\vert 0\right\rangle =D_{\mu\nu,\alpha\beta}(x-y),  \label{eq:prop1}
\end{equation}
where $D_{\mu\nu,\alpha\beta}$ is the operator that satisfies the Green'{}s
equation, given as
\begin{equation}
\hat{\mathcal{O}}_{\ \ \lambda\sigma}^{\mu\nu,}D^{\lambda\sigma,\alpha\beta
}(x-y)=i\mathcal{I}^{\mu\nu,\alpha\beta}\delta^{4}(x-y),
\end{equation}
with $\mathcal{I}^{\mu\nu,\alpha\beta}=\frac{1}{2}\left(
\eta^{\mu\alpha}\eta^{\nu\beta}+\eta^{\mu\beta}\eta^{\nu\alpha}\right) $
playing the role of the identity operator. Thus, the problem of determining
the propagator is reduced to the inversion of the operator $\hat{\mathcal{O}}
$, given in Eq. (\ref{eq:Operator}). Once found a closed operator algebra
composed of a set of appropriated tensor projectors\ with which the operator
$\hat{\mathcal{O}}$ can be expanded, the inversion of the operator becomes a
tedious but straightforward task.

As it is well known, a convenient method for obtaining the inverse of
symmetric rank-two tensors is based on the spin projector operators found by
Barnes and Rivers \cite{Barnes,Rivers,Sezgin,Boldo} which constitute a
complete orthonormal basis of operators for Lorentz-invariant models in four
dimensions. This basis is shown in Eq. \eqref{eq:P1}. To accommodate the
emerging terms containing the LV background vector $b^{\mu}$ in gravity
theories, an extended basis of the Barnes-Rivers projectors was devised in
Ref. \cite{Boldo}. All tools needed to invert our operator $\hat{\mathcal{O}}
$ are outlined in the Appendix.

Using the spin-projection operators and the identities as given in the
Appendix \ref{sec:Appendix-A:-BarnesRiversOperators}, we are able to put the
operator $\hat{\mathcal{O}}$ in the form (where for simplicity we adopt the
notation $AB$ in place of $A^{\mu\nu}{}_{\rho\sigma}B^{\rho\sigma\alpha%
\beta} $ to the contractions)
\begin{equation}
\hat{\mathcal{O}}=a_{1}\text{P}^{(1)}+a_{2}\text{P}^{(2)}+a_{3}\text{P}%
^{(0-\theta)}+a_{4}\text{P}^{(0-\theta\omega)}+a_{5}{\tilde{\Pi}}^{(1)}+a_{6}%
\tilde{\Pi}^{(2)}+a_{7}{\tilde{\Pi}}^{(\theta\Sigma)}+a_{8}{\tilde{\Pi}}%
^{(\theta\Lambda)},  \label{ope1}
\end{equation}
with the scalar coefficients $a_{i}$ being functions of the momentum and the
background vector $b^{\mu}$ given explicitly by
\begin{align}
a_{1} & =-\frac{4\xi^{2}(b\cdot p)^{2}}{\kappa^{2}}+\xi(b\cdot
p)^{2},~~a_{2}=\xi(b\cdot p)^{2}+p^{2},~  \notag \\[0.1cm]
a_{3} & =\frac{24\xi^{2}\boxdot(p)}{\kappa^{2}}-\frac{6\xi^{2}p^{4}}{%
\kappa^{2}\lambda}-2\xi(b\cdot p)^{2}-2p^{2},~a_{5}=\frac{4\xi^{2}\left(
b\cdot p\right) }{\kappa^{2}},  \notag \\[0.1cm]
a_{4} & =\frac{8\sqrt{3}\xi^{2}(b\cdot p)^{2}}{\kappa^{2}}-\sqrt{3}%
\xi(b\cdot p)^{2},\ a_{6}=-\frac{4\xi^{2}p^{2}}{\kappa^{2}}-\xi p^{2},~\  \\%
[0.1cm]
~~a_{7} & =-\frac{8\sqrt{3}\xi^{2}\left( b\cdot p\right) }{\kappa^{2}}%
,~a_{8}=\frac{8\sqrt{3}\xi^{2}p^{2}}{\kappa^{2}}-\sqrt{3}\xi p^{2}.  \notag
\end{align}

Note that the projectors $\tilde{\Pi},$\ defined in the Appendix \ref{sec:Appendix-A:-BarnesRiversOperators}, are the
ones carrying the bumblebee field. The upper tilde symbol\ is used to
highlight the difference in relation to the projectors first defined in
Ref. \cite{Boldo}.

After a lengthy computation, and using the identity $\hat {%
\mathcal{O}}\hat{\mathcal{O}}^{-1}=I$, we are able to write the
operator $\hat{\mathcal{O}}^{-1}$ in terms of the whole set of projectors,
as
\begin{align}
\hat{\mathcal{O}}^{-1} & =b_{1}\text{P}^{(1)}+b_{2}\text{P}^{(2)}+b_{3}\text{%
P}^{(0-\theta)}+b_{4}\text{P}^{(0-\omega)}+b_{5}\text{{P}}%
^{(0-\theta\omega)}+b_{6}{\tilde{\Pi}}^{(1)}+b_{7}\tilde{\Pi}^{(2)}+b_{8}{%
\tilde{\Pi}}^{(\theta\Sigma)}  \notag \\[0.3cm]
& +b_{9}{\tilde{\Pi}}^{(\theta\Lambda)}+b_{10}\tilde{\Pi}^{(\Lambda\Lambda
)}+b_{11}{\tilde{\Pi}}^{(\omega\Lambda-a)}+b_{12}{\tilde{\Pi}}^{(\omega
\Lambda-b)}+b_{13}{\tilde{\Pi}}^{(\omega\Sigma)}+b_{14}{\tilde{\Pi}}%
^{(\Lambda\Sigma)},
\end{align}
where the coefficients $b_{i}$\ are constrained by a set of fourteen
algebraic equations. Solving them, we obtain the coefficients
\begin{align}
b_{1} & =\frac{N_{1}}{\kappa^{2}\xi^{2}(b\cdot p)^{2}\boxdot\boxplus},~b_{2}=%
\frac{1}{\boxplus},~b_{3}=-\frac{1}{2\boxplus},~b_{4}=\frac{N_{4}}{%
2\lambda\kappa^{2}\xi^{2}(b\cdot p)^{4}\boxdot^{2}\boxplus},~  \notag \\%
[0.1cm]
~\ b_{5} & =\frac{N_{5}}{2\xi(b\cdot p)^{2}\boxdot\boxplus},~b_{6}=\frac{%
p^{2}}{\xi(b\cdot p)\boxdot\boxplus},~b_{7}=\frac{p^{2}}{\boxdot \boxplus},
~b_{8}=\frac{N_{8}}{4\xi(b\cdot p)\boxdot\boxplus},~ \\[0.1cm]
b_{9} & =-\frac{\sqrt{3}p^{2}}{2\boxdot\boxplus},~b_{10}=\frac{p^{4}}{%
2\boxdot^{2}\boxplus},~b_{11}=\frac{N_{11}}{8\xi^{2}(b\cdot
p)^{2}\boxdot^{2}\boxplus},~b_{12}=\frac{N_{12}}{2\xi(b\cdot p)^{2}\boxdot
^{2}\boxplus},~  \notag \\
b_{13} & =\frac{N_{13}}{4\kappa^{2}\xi^{2}(b\cdot p)^{3}\boxdot^{2}\boxplus }%
,~b_{14}=\frac{N_{14}}{4\xi\left( b\cdot p\right) \boxdot^{2}\boxplus },
\notag
\end{align}
where one has used $\boxplus=\boxplus(p)$\ and $\boxdot=\boxdot(p),$\
defined by the following expressions\textbf{\ }%
\begin{align}
\boxplus(p) & =p^{2}+\xi(b\cdot p)^{2}, \\[0.1cm]
\boxdot(p) & =(b\cdot p)^{2}-b^{2}p^{2}.
\end{align}
Furthermore, the numerators $N_{i}$\ follow
\begin{align}
N_{1} & =\xi(4\xi+\kappa^{2})\boxdot\boxplus+\kappa^{2}p^{4}, \\[0.1cm]
N_{4} & =\xi^{2}\boxdot^{2}\left[ p^{2}F_{1}\left( p\right) +\lambda
\kappa^{2}(b\cdot p)^{4}\right] +4\lambda\xi\kappa^{2}p^{2}(b\cdot
p)^{4}\boxdot \\[0.1cm]
& +\xi^{3}(b\cdot p)^{2}\boxdot^{2}F_{1}\left( p\right) -\lambda\kappa
^{2}p^{4}\left[ b^{4}p^{4}-4(b\cdot p)^{4}+2b^{2}p^{2}(b\cdot p)^{2}\right] ,
\notag \\[0.1cm]
N_{5} & =\sqrt{3}\left[ -\xi(b\cdot p)^{2}\boxdot+b^{2}p^{4}\right] , ~
N_{8} =\sqrt{3}\left( \xi(b\cdot p)^{2}-p^{2}\right) , \\[0.1cm]
N_{11} & =p^{2}\left( p^{2}-\xi(b\cdot p)^{2}\right) ^{2},\text{ }%
N_{12}=p^{4}\left( 2b^{2}p^{2}-3(b\cdot p)^{2}\right) -2\xi p^{2}(b\cdot
p)^{2}\boxdot,~ \\[0.1cm]
N_{13} & =\ F_{2}\left( p\right) \xi^{2}\boxdot-16\xi^{3}(b\cdot
p)^{2}\boxdot^{2}+\xi\kappa^{2}b^{2}p^{4}\left[ 2(b\cdot p)^{2}-b^{2}p^{2}%
\right] \\[0.1cm]
& +\kappa^{2}p^{4}\left[ 2b^{2}p^{2}-3(b\cdot p)^{2}\right] ,  \notag \\%
[0.1cm]
N_{14} & =p^{2}\left[ p^{2}-\xi(b\cdot p)^{2}\right] ,
\end{align}
and
\begin{align}
F_{1}\left( p\right) & =16\lambda(b\cdot p)^{2}+16\lambda b^{2}p^{2}+p^{4},
\\[0.1cm]
F_{2}\left( p\right) & =\kappa^{2}(b\cdot p)^{2}\left[ (b\cdot
p)^{2}+b^{2}p^{2}\right] +16p^{2}\left[ b^{2}p^{2}-(b\cdot p)^{2}\right] .
\end{align}

The Feynman propagator is
\begin{align}
D_{\mu\nu,\alpha\beta}(p) & =\frac{i}{\boxplus(p)}\left\{ \frac{N_{1}}{%
\kappa^{2}\xi^{2}(b\cdot p)^{2}\boxdot}\text{P}_{\mu\nu,\alpha\beta}^{(1)}+%
\text{P}_{\mu\nu,\alpha\beta}^{(2)}-\frac{1}{2}\text{P}_{\mu\nu
,\alpha\beta}^{(0-\theta)}+\frac{N_{4}}{2\lambda\kappa^{2}\xi^{2}(b\cdot
p)^{2}\boxdot^{2}}\text{P}_{\mu\nu,\alpha\beta}^{(0-\omega)}\right.  \notag
\\[0.15cm]
& +\frac{p^{2}}{\boxdot}{\tilde{\Pi}}_{\mu\nu,\alpha\beta}^{(2)}+\frac{N_{5}%
}{2\xi(b\cdot p)^{2}\boxdot}\text{P}_{\mu\nu,\alpha\beta}^{(0-\theta\omega
)}+\frac{p^{2}}{\xi(b\cdot p)\boxdot}{\tilde{\Pi}}_{\mu\nu,\alpha%
\beta}^{(1)}+\frac{N_{8}}{4\xi(b\cdot p)\boxdot}{\tilde{\Pi}}%
_{\mu\nu,\alpha\beta }^{(\theta\Sigma)}  \notag \\[0.15cm]
& -\frac{\sqrt{3}p^{2}}{2\boxdot}{\tilde{\Pi}}_{\mu\nu,\alpha\beta}^{(\theta%
\Lambda)}+\frac{p^{4}}{2\boxdot^{2}}{\tilde{\Pi}}_{\mu\nu,\alpha
\beta}^{(\Lambda\Lambda)}+\frac{N_{11}}{8\xi^{2}(b\cdot p)^{2}\boxdot^{2}}{%
\tilde{\Pi}}_{\mu\nu,\alpha\beta}^{(\omega\Lambda-a)}  \label{prop_grav} \\%
[0.15cm]
& \left. +\frac{N_{12}}{2\xi(b\cdot p)^{2}\boxdot^{2}}{\tilde{\Pi}}_{\mu
\nu,\alpha\beta}^{(\omega\Lambda-b)}+\frac{N_{13}}{4\kappa^{2}\xi^{2}(b\cdot
p)^{3}\boxdot^{2}}{\tilde{\Pi}}_{\mu\nu,\alpha\beta}^{(\omega\Sigma)}+\frac{%
N_{14}}{4\xi(b\cdot p)\boxdot^{2}}{\tilde{\Pi}}_{\mu\nu,\alpha\beta
}^{(\Lambda\Sigma)}\right\} .  \notag
\end{align}

The next step is to read off the graviton dispersion relations from the
poles of the propagator, aiming at verifying the consistency of the theory
concerning causality and unitarity respects.

\section{Dispersion relations\label{sec:DR}}

In this section, we analyze the dispersion relations stemming from the poles
of the graviton propagator, which provide information about the stability
and causality of the modes.

We begin considering the pole $\boxplus(p),$\ which implies
\begin{equation}
p^{2}+\xi(b\cdot p)^{2}=0.  \label{DR1}
\end{equation}
For the timelike configuration, $b^{\mu}=\left( b_{0},\mathbf{0}\right) ,$\
the corresponding dispersion relation is
\begin{equation}
p_{0}=\pm\frac{\left\vert \mathbf{p}\right\vert }{\sqrt{1+\xi b_{0}^{2}}},
\label{DR1a}
\end{equation}
which is a positive energy mode and yields the group velocity
\begin{equation}
u_{g}=\frac{1}{\sqrt{1+\xi b_{0}^{2}}},
\end{equation}
which is smaller than 1 for $\xi>0,$ implying causality assurance for $\xi>0.$

For the spacelike configuration, $b^{\mu}=\left( 0,\mathbf{b}\right) ,$\ the
dispersion relation is\textbf{\ }%
\begin{equation}
p_{0}=\pm\left\vert \mathbf{p}\right\vert \sqrt{1-\xi\left\vert \mathbf{b}%
\right\vert ^{2}\cos^{2}\theta},  \label{DR1b}
\end{equation}
where $(b\cdot p)=\left\vert \mathbf{b}\right\vert \left\vert \mathbf{p}%
\right\vert \cos\theta.$\ This is a positive energy mode, related with the
group velocity\
\begin{equation}
u_{g}=\sqrt{1-\xi\left\vert \mathbf{b}\right\vert ^{2}\cos^{2}\theta },%
\mathbf{\ }
\end{equation}
which becomes\ smaller than $1$\ for $\xi>0$\ and $\xi\left\vert \mathbf{b}%
\right\vert ^{2}<1$. So, this mode is causal for $\xi>0$\ for both
configurations.

For the pole $\boxdot (p)$, the associated dispersion relation is given by
the roots of%
\begin{equation}
(b\cdot p)^{2}-b^{2}p^{2}=0.  \label{DR2}
\end{equation}%
In a general background, $b^{\mu }=\left( b_{0},\mathbf{b}\right) $, the
dispersion relation is
\begin{equation}
p_{0}=\frac{\left\vert \mathbf{p}\right\vert }{\left\vert \mathbf{b}%
\right\vert }\left[ b_{0}\cos \theta \pm \sqrt{\left( \left\vert \mathbf{b}%
\right\vert ^{2}-b_{0}^{2}\right) \sin ^{2}\theta }\right] ,  \label{DR2b}
\end{equation}%
It becomes clear that the condition $\left\vert \mathbf{b}%
\right\vert ^{2}>b_{0}^{2}$ ensures the existence of real roots. 
Hence, the background $b^{\mu }$ must be
spacelike. In order to facilitate the analysis, we adopt the simplest
spacelike background, $b^{\mu }=(0,\mathbf{b}),$ which is
equivalent to any other spacelike choice due to the observer Lorentz
symmetry.\ For this background, the dispersion relation appears as 
\begin{equation}
p_{0}=\pm \left\vert \mathbf{p}\right\vert \sin \theta ,  \label{DR2b1}
\end{equation}%
providing a causal mode, $u_{g}=\sin \theta \leq 1$, whose
energy presents a strong dependence on the direction of propagation. Besides
this physically unusual behavior, the relation \eqref{DR2b1} yields a
nonunitary mode, as it will be shown in the next section.


\section{Tree-level unitarity}

The tree-level unitarity analysis of this model is performed through the
saturation of the Feynman propagator with external currents. This method is
usually applied\ in quantum field theory \cite{Veltman}, being implemented
by means of the saturated propagator\ ($SP$), $SP=J^{\ast\mu}$Res$%
(\Delta_{\mu \nu})$\ $J^{\nu},$ a scalar quantity given by the contraction
of the external currents $(J^{\mu})$\ with Res$(\Delta_{\mu\nu})$ -\ the
residue of the propagator evaluated at each pole. The conserved current$%
,\partial_{\mu}J^{\mu}=0,$\ implies $p_{\mu}J^{\mu}=0.$\textbf{\ }This
method was already used to analyze unitarity in the gauge sector of the SME
\cite{Casana1}.

For the rank-two graviton field, this method can also be equally applied. In
this case the saturated residue of propagator is written as
\begin{equation}
SP=J^{\mu\nu}\left( \text{Res}D_{\mu\nu,\kappa\lambda}\right) J^{\kappa
\lambda},
\end{equation}
where $\left( \text{Res}D_{\mu\nu,\kappa\lambda}\right) $\ is the residue
evaluated at each pole of the propagator, and $J^{\mu\nu}$\ is a symmetric
tensor describing an external conserved current\ $\left(
\partial_{\mu}J^{\mu\nu}=0\right) $, which in momentum space reads as $%
p_{\mu}J^{\mu\nu}=0$. In accordance with this method, the unitarity analysis
is assured whenever the imaginary part of the saturation $SP$\ (at the poles
of the propagator) is positive.

Because of the conservation law, $p_{\mu}J^{\mu\nu}=0,$\ all the projector terms
involving $\omega_{\mu\nu}$\ and $\Sigma_{\mu\nu}$\ yield null saturation.
Hence, non-null contribution for saturation stems only from the following
terms:
\begin{align}
J^{\mu\nu}\text{P}^{(2)}{}_{\mu\nu,\kappa\lambda}J^{\kappa\lambda} &
=J_{\kappa\lambda}J^{\kappa\lambda}-\frac{1}{3}\left( J^{\kappa}{}_{\kappa
}\right) ^{2},  \notag \\[0.1cm]
J^{\mu\nu}\text{P}^{(0-\theta)}{}_{\mu\nu,\kappa\lambda}J^{\kappa\lambda} & =%
\frac{1}{3}\left( J^{\kappa}{}_{\kappa}\right) ^{2},  \notag \\[0.1cm]
J^{\mu\nu}\tilde\Pi_{\mu\nu,\kappa\lambda}^{(2)}J^{\kappa\lambda} &
=2b_{\nu}J^{\nu}{}_{\kappa}b_{\lambda}J^{\lambda\kappa}~, \\[0.1cm]
J^{\mu\nu}\tilde\Pi^{(\theta\Lambda)}{}_{\mu\nu,\kappa\lambda}J^{\kappa%
\lambda} & =\frac{2}{\sqrt{3}}J^{\kappa}{}_{\kappa}b_{\mu}b_{\nu}J^{\mu\nu },
\notag \\[0.1cm]
J^{\mu\nu}\tilde\Pi_{\mu\nu,\kappa\lambda}^{(\Lambda\Lambda)}J^{\kappa%
\lambda} & =J^{\mu\nu}b_{\mu}b_{\nu}b_{\kappa}b_{\lambda}J^{\kappa\lambda}.
\notag
\end{align}
Using the Feynman propagator (\ref{prop_grav}) and the current conservation,
the propagator saturated by conserved currents reads
\begin{equation}
SP=\frac{J_{\kappa\lambda}J^{\kappa\lambda}-\frac{1}{2}\left(
J^{\kappa}{}_{\kappa}\right) ^{2}}{\boxplus}+\frac{2p^{2}b_{\nu}J^{\nu}{}_{%
\kappa }b_{\lambda}J^{\lambda\kappa}}{\boxplus\boxdot}-\frac{%
p^{2}J^{\kappa}{}_{\kappa}b_{\mu}b_{\nu}J^{\mu\nu}}{\boxplus\boxdot}+\frac{%
p^{4}J^{\mu\nu }b_{\mu}b_{\nu}b_{\kappa}b_{\lambda}J^{\kappa\lambda}}{%
2\boxplus\boxdot^{2}}.  \label{satured}
\end{equation}
Next, we compute the residues in the poles of the propagator, whose
corresponding dispersion relations were studied in Sec. \ref{sec:DR}.

\subsection{The first pole $\boxplus=p^{2}+\protect\xi(b\cdot p)^{2}$}

This pole implies the dispersion relation (\ref{DR1}). The corresponding
residue obtained from (\ref{satured}) yields the following expression:%
\begin{equation}
\hspace{-0.75cm}\left. \frac{{}}{{}}\text{Res}\left( S\right) \right\vert
_{\boxplus=0}\!\!=\!\!J_{\kappa\lambda}J^{\kappa\lambda}-\frac{1}{2}\left(
J^{\kappa}{}_{\kappa}\right) ^{2}-\frac{2\xi\left( b_{\nu}J^{\nu}{}_{\kappa
}\right) ^{2}}{\left( 1+\xi b^{2}\right) }+\frac{\xi\left(
J^{\kappa}{}_{\kappa}\right) \left( b_{\mu}b_{\nu}J^{\mu\nu}\right) }{\left(
1+\xi b^{2}\right) }+\frac{\xi^{2}\left( b_{\mu}b_{\nu}J^{\mu\nu}\right) ^{2}%
}{2\left( 1+\xi b^{2}\right) ^{2}},  \label{Res1}
\end{equation}
where $b^{2}=b^{\mu}b_{\mu}$.

In the following calculations we will use the relations%
\begin{equation}
J_{00}=\frac{p_{a}p_{c}}{p_{0}^{2}}J_{ca},~\ \ J_{0a}=\frac{p_{c}}{p_{0}}%
J_{ca},  \label{RRes_c}
\end{equation}
obtained from the current conservation condition.

We should now specialize our analysis for two cases: a timelike background, $%
b^{\mu}=(b_{0},\mathbf{0}),$\ and a spacelike background, $b^{\mu }=(0,%
\mathbf{b})$.

For the timelike background, $b^{\mu}=(b_{0},\mathbf{0}),$\ the dispersion
relation is written as Eq. (\ref{DR1a}), and the residue (\ref{Res1})
becomes
\begin{equation}
\!\!\!\!\left. \frac{{}}{{}}\text{Res}\left( S\right) \right\vert
_{b=(b_{0},0)}\!=\!\frac{1}{2}\left[ \frac{p_{a}p_{c}J_{ca}}{\mathbf{p}^{2}}%
+\left( J_{dd}\right) \right] ^{2}+\left( J_{ab}\right) ^{2}-\left(
J_{dd}\right) ^{2}-2\frac{\left( p_{c}J_{ca}\right) ^{2}}{\mathbf{p}^{2}}.
\label{res_pole1}
\end{equation}
It is exactly equal to the one stemming from the usual graviton mode $%
p^{2}=0 $\ of the Einstein-Hilbert's gravity which preserves unitarity.
Therefore, the residue (\ref{res_pole1}) is positive definite and the pole $%
p^{2}+\xi(b\cdot p)^{2}=0$, is unitary for all values of $b_{0}.$

For the spacelike background, $b^{\mu }=(0,\mathbf{b}),$ the dispersion
relation is given by Eq. (\ref{DR1b}), or
\begin{equation}
\left( p_{0}\right) ^{2}=\mathbf{p}^{2}-\xi \left( \mathbf{b\cdot p}\right)
^{2}.
\end{equation}%
The residue (\ref{Res1}) for this background configuration is
evaluated by using the following set of orthogonal vectors%
\begin{equation}
\mathbf{u}_{1}=\mathbf{u}_{3}\times \mathbf{u}_{b},~\ \ \mathbf{u}_{2}=%
\mathbf{u}_{3}\times \mathbf{u}_{1},~\ \ \mathbf{u}_{3}=\mathbf{p/}%
\left\vert \mathbf{p}\right\vert ,
\end{equation}%
where$\ \mathbf{u}_{b}=\mathbf{b}/\left\vert \mathbf{b}\right\vert $\ has
the direction of the LV background. So\ the residue (\ref{Res1}) results are 
\begin{equation}
\left. \!\!\!\frac{{}}{{}}\text{Res}\left( S\right) \right\vert _{b=(0,%
\mathbf{b})}\!\!\!\!=\!\!\frac{1}{1-\xi \mathbf{b}^{2}}\left[ \frac{2\left(
S_{1}\right) ^{2}}{1-\xi \left( b_{3}\right) ^{2}}+\frac{\left( S_{2}\right)
^{2}}{2\left( 1-\xi \mathbf{b}^{2}\right) }\right] ,  \label{RESS_2}
\end{equation}%
with the terms
\begin{equation*}
S_{1}=\left[ 1-\xi \left( b_{3}\right) ^{2}\right] J_{12}^{u}+\xi
b_{2}b_{3}J_{13}^{u},
\end{equation*}%
\begin{equation}
S_{2}=\frac{\left( \xi b_{2}b_{3}\right) ^{2}J_{33}^{u}}{1-\xi \left(
b_{3}\right) ^{2}}+J_{22}^{u}\left[ 1-\xi \left( b_{3}\right) ^{2}\right]
-J_{11}^{u}\left( 1-\xi \mathbf{b}^{2}\right) +2\xi b_{2}b_{3}J_{23}^{u}.
\end{equation}%
Here, we have used the definitions $\mathbf{b=}b_{3}\mathbf{u}_{3}+b_{2}%
\mathbf{u}_{2}$\textbf{, }and $J_{ij}^{u}=\mathbf{u}_{i}\cdot \left( \mathbb{%
J}\mathbf{u}_{j}\right) $\textbf{, }with $\mathbb{J}=\left[ J_{ij}\right] $%
\textbf{\ \ }and $J_{ij}^{u}$ being the spatial components of the tensor $%
J_{\mu \nu }$ in the basis $\left\{ \mathbf{u}_{1},\mathbf{u}_{2},\mathbf{u}%
_{3}\right\} $\textbf{. }The residue (\ref{RESS_2}) is positive whenever
\begin{equation}
\xi \mathbf{b}^{2}<1.
\end{equation}%
As the magnitude of the LV background should be small, the unitary of this
mode, for $b^{\mu }=(0,\mathbf{b}),$\ is assured. Hence, the pole $%
\boxplus$ provides causal and unitary propagating modes.

\subsection{The second pole $\boxdot=(b\cdot p)^{2}-b^{2}p^{2}$}

It is a double pole\ implying the dispersion relation (\ref{DR2b}),\ which
is physically sensible for $b_{0}=0,$ that is, a spacelike
background, $b^{\mu }=(0,\mathbf{b}).$ Its residue, computed from the
saturated propagator (\ref{satured}) is%
\begin{equation}
\left. \frac{{}}{{}}\text{Res}\left( S\right) \right\vert _{\boxdot
=0}=R_{1}+R_{2},
\end{equation}%
with%
\begin{align}
R_{1}& =-\frac{2b_{\nu }J^{\nu }{}_{\kappa }b_{\lambda }J^{\lambda \kappa
}-J^{\kappa }{}_{\kappa }b_{\mu }b_{\nu }J^{\mu \nu }}{b^{2}\left( 1+\xi
b^{2}\right) },  \label{r1} \\
~\ R_{2}& =\frac{\left( J^{\mu \nu }b_{\mu }b_{\nu }\right) ^{2}}{%
2b^{4}\left( 1+\xi b^{2}\right) }\frac{\left( 1+2\xi b^{2}\right) }{\left(
1+\xi b^{2}\right) }.  \label{r2}
\end{align}%
Using the identities (\ref{RRes_c}), the quantities $J^{\mu \nu }b_{\mu
}b_{\nu }$, $\ b_{\nu }J^{\nu }{}_{\kappa }b_{\lambda }J^{\lambda \kappa }~$%
and $\ J^{\kappa }{}_{\kappa }$ become
\begin{equation}
J^{\mu \nu }b_{\mu }b_{\nu }=\frac{\left( b_{0}\right) ^{2}\left( \mathbf{%
p\cdot }\mathbb{J}\mathbf{p}\right) }{p_{0}^{2}}-2\frac{b_{0}\left( \mathbf{%
b\cdot }\mathbb{J}\mathbf{p}\right) }{p_{0}}+\mathbf{b\cdot }\mathbb{J}%
\mathbf{b},  \label{Jbb}
\end{equation}%
\begin{align}
b_{\nu }J^{\nu }{}_{\kappa }b_{\lambda }J^{\lambda \kappa }& =\frac{\left(
b_{0}\right) ^{2}\left( \mathbf{p\cdot }\mathbb{J}\mathbf{p}\right) ^{2}}{%
\left( p_{0}\right) ^{4}}-2\frac{b_{0}\left( \mathbf{p\cdot }\mathbb{J}%
\mathbf{p}\right) \left( \mathbf{b\cdot }\mathbb{J}\mathbf{p}\right) }{%
\left( p_{0}\right) ^{3}}+\frac{\left( \mathbf{b\cdot }\mathbb{J}\mathbf{p}%
\right) ^{2}}{\left( p_{0}\right) ^{2}}  \label{Jbb_1} \\
& -\frac{\left( b_{0}\right) ^{2}\left( \mathbf{p\cdot }\mathbb{J}^{2}%
\mathbf{p}\right) }{\left( p_{0}\right) ^{2}}+2\frac{b_{0}\left( \mathbf{%
b\cdot }\mathbb{J}^{2}\mathbf{p}\right) }{p_{0}}-\mathbf{b\cdot }\mathbb{J}%
^{2}\mathbf{b},  \notag
\end{align}%
\begin{equation}
J^{\kappa }{}_{\kappa }=J_{00}-J_{aa}=\frac{\left( \mathbf{p\cdot }\mathbb{J}%
\mathbf{p}\right) }{p_{0}^{2}}-J_{aa},  \label{Jbb_2}
\end{equation}%
\textbf{where} $\left( \mathbf{a\cdot }\mathbb{J}\mathbf{b}\right) =$ $%
J^{lm}a_{l}b_{m},$ $\left( \mathbf{a\cdot }\mathbb{J}^{2}\mathbf{p}\right) =$
$J^{ln}J^{nm}a_{l}p_{m}.$ We now analyze this pole in two cases.

For $b_{0}=0$, the dispersion relation is
\begin{equation*}
p_{0}=\pm \frac{\left\vert \mathbf{p\times b}\right\vert }{\left\vert
\mathbf{b}\right\vert },
\end{equation*}%
and the expressions (\ref{r1})--(\ref{r2}) can be written as
\begin{align}
\mathbf{b}^{2}\left( 1-\xi \mathbf{b}^{2}\right) R_{1}& =\frac{2\left(
\mathbf{p\cdot }\mathbb{J}\mathbf{b}\right) ^{2}}{\left( p_{0}\right) ^{2}}%
-2\left( \mathbf{b\cdot }\mathbb{J}^{2}\mathbf{b}\right) -\frac{\left(
\mathbf{b\cdot }\mathbb{J}\mathbf{b}\right) \left( \mathbf{p\cdot }\mathbb{J}%
\mathbf{p}\right) }{\left( p_{0}\right) ^{2}}+\left( \mathbf{b\cdot }\mathbb{%
J}\mathbf{b}\right) J_{aa}, \\[0.2cm]
\mathbf{b}^{2}\left( 1-\xi \mathbf{b}^{2}\right) R_{2}& =\frac{\left(
\mathbf{b\cdot }\mathbb{J}\mathbf{b}\right) ^{2}}{2\mathbf{b}^{2}}\frac{%
\left( 1-2\xi \mathbf{b}^{2}\right) }{\left( 1-\xi \mathbf{b}^{2}\right) }.
\end{align}%
To simplify the above expressions, we use the orthonormal basis
\begin{equation}
\mathbf{u}_{1}=\frac{\mathbf{p\times b}}{\left\vert \mathbf{p\times b}%
\right\vert }=\frac{\mathbf{p\times u}_{3}}{p_{0}},~\mathbf{u}_{2}=\mathbf{u}%
_{3}\times \mathbf{u}_{1},~\mathbf{u}_{3}=\frac{\mathbf{b}}{\left\vert
\mathbf{b}\right\vert },
\end{equation}%
so we have the following expressions: $\mathbf{p}=p_{2}\mathbf{u}_{2}+p_{3}%
\mathbf{u}_{3},~\mathbf{b}=\left\vert \mathbf{b}\right\vert \mathbf{u}_{3},~%
\mathbf{p\times b=}\left\vert \mathbf{b}\right\vert p_{2}\mathbf{u}_{1}$, $%
J_{ij}^{u}=\mathbf{u}_{i}\cdot \left( \mathbb{J}\mathbf{u}_{j}\right) $%
\textbf{, }with $\mathbb{J}=\left[ J_{ij}\right] $ being $J_{ij}$\ the
spatial components of the tensor $J_{\mu \nu }$\textbf{. }After some algebra
we obtain the residue%
\begin{equation}
\left. \frac{{}}{{}}\text{Res}\left( S\right) \right\vert _{\boxdot =0}=%
\frac{J_{11}^{u}J_{33}^{u}-2\left( J_{13}^{u}\right) ^{2}}{1-\xi \mathbf{b}%
^{2}}+\frac{2p_{2}p_{3}J_{23}^{u}J_{33}^{u}+\left( p_{3}\right) ^{2}\left(
J_{33}^{u}\right) ^{2}}{\left( p_{2}\right) ^{2}\left( 1-\xi \mathbf{b}%
^{2}\right) }-\frac{\left( J_{33}^{u}\right) ^{2}}{2\left( 1-\xi \mathbf{b}%
^{2}\right) ^{2}},  \label{Res_end2}
\end{equation}%
which is not definite positive, meaning, in general, a nonunitary
excitation.

It means that this double pole provides excitations that
spoil the unitarity of the model (ghost excitations) and, consequently, its
physical consistency. The spoiling role can be, in principle, ascribed to
the tensor operators $\tilde{\Pi}_{\mu \nu ,\kappa \lambda }^{(2)},$ $%
\tilde{\Pi}^{(\theta \Lambda )}{}_{\mu \nu ,\kappa \lambda },$ $\tilde{\Pi}%
_{\mu \nu ,\kappa \lambda }^{(\Lambda \Lambda )},$ which contribute
to the scalar saturation (\ref{satured}). More specifically, the projector $%
\tilde{\Pi}_{\mu \nu ,\kappa \lambda }^{(\Lambda \Lambda )}$ is the
responsible for the unitarity breaking, since it is the unique one
associated with a second-order pole, a feature generally related to ghost
excitations in quantum field theory.

Therefore, this gravity model despite have a simples pole (\ref{DR1}) which preserves the tree-level unitarity,  is nonunitary due to the presence of a second-order pole.

\section{Degrees of freedom}

Now, we show that the gravitational field modified by the vacuum expectation
value of the bumblebee field is still represented for a massless spin two
symmetric tensor field with only two physical degrees of freedom. The
equation of motion obtained from the effective Lagrangian (\ref{L_E}) is
\begin{equation}
\hat{\mathcal{O}}_{\mu\nu,\alpha\beta}h^{\alpha\beta}=0,  \label{eq:EOM}
\end{equation}
with the operator $\hat{\mathcal{O}}$\ defined by Eq. (\ref{ope1}). By
saturating Eq. (\ref{eq:EOM}) with $p^{\mu}p^{\nu}$, we obtain the following
constraint:
\begin{equation}
p_{\mu}p_{\nu}h^{\mu\nu}=p^{2}h,  \label{eq:constrain1}
\end{equation}
where we have assumed $\xi\neq0$\ and $(b\cdot p)\neq0$. As we can easily
observe, due to the presence of the background field $b_{\mu}$, we can still
saturate Eq. (\ref{eq:EOM}) with the combinations $b^{(\mu}p^{\nu)}$\ and $%
b^{\mu}b^{\nu}$. In addition we also perform the saturation of the equation
of motion (\ref{eq:EOM}) with the metric $\eta_{\mu\nu}$. Such procedures
imply the following equations,\
\begin{align}
0 & =(b\cdot p)^{2}h-2(b\cdot
p)b_{(\mu}p_{\nu)}h^{\mu\nu}+p^{2}b_{\mu}b_{\nu}h^{\mu\nu}, \\[0.2cm]
0 & =\left[ p^{2}\left( 1-3\xi b^{2}\right) +2\xi(b\cdot p)^{2}\right]
b_{\mu}b_{\nu}h^{\mu\nu}+(b\cdot p)^{2}\left( 1-\xi b^{2}\right) h-2(b\cdotp%
)\left( 1-\xi b^{2}\right) b_{(\mu}p_{\nu)}h^{\mu\nu },  \notag \\[0.2cm]
0 & =-2\left( b\cdot p\right) \left( 16\xi-\kappa^{2}\right) b_{(\mu
}p_{\nu)}h^{\mu\nu}+p^{2}\left( 16\xi-5\kappa^{2}\right) b_{\mu}b_{\nu
}h^{\mu\nu}+(b\cdot p)^{2}\left( 16\xi-3\kappa^{2}\right) h,  \notag
\end{align}
where we have used the condition (\ref{eq:constrain1}). Thus, it is
straightforward to see that these three equations imply the following
constraint relations which $h_{\mu\nu}$\ and $b_{\mu}$\ must satisfy:
\begin{align}
b_{\mu}b_{\nu}h^{\mu\nu} & =0,  \label{eq:v1} \\
b_{(\mu}p_{\nu)}h^{\mu\nu} & =0,  \label{eq:v2} \\
h & =0.  \label{eq:v3}
\end{align}
Immediately, from (\ref{eq:constrain1}) it follows that
\begin{equation}
p_{\mu}p_{\nu}h^{\mu\nu}=0.  \label{eq:v4}
\end{equation}
We can achieve even more restrictions when we perform the contraction of Eq.
(\ref{eq:EOM}) with $p_{\mu}$\ or $b_{\mu}$\ separately. By following a
similar procedure as above, we\ obtain the following constraints:
\begin{align}
p_{\mu}h_{\ \nu}^{\mu} & =0,  \label{eq:vin5} \\
b_{\mu}h_{\ \nu}^{\mu} & =0,  \label{eq:vin6}
\end{align}
representing a total of 8 additional conditions to be satisfied by the
fields. Thus, the set of Eqs. (\ref{eq:v1})-(\ref{eq:vin6}) provides a
total of 12 constraints which can be used to reduce the 14 initial degrees
of freedom contained in the graviton and bumblebee fields. Consequently,
we are left with only two physical degrees of freedom such as it happens for
the Einstein-Hilbert's graviton.

Finally, by applying Eqs. (\ref{eq:v1})--(\ref{eq:vin6}) to Eq. (\ref{eq:EOM}%
),\ it becomes simply
\begin{equation}
\left[ p^{2}+\xi (b\cdot p)^{2}\right] h_{\mu \nu }=0,
\end{equation}%
providing the correct energy-momentum dispersion relation (\ref{DR1})\
associated to the physical pole as previously determined. Therefore, we can
conclude that the mechanism of spontaneous Lorentz-symmetry breaking
triggered by the bumblebee field has modified the Einstein-Hilbert dispersion
relation as: $p^{2}=0\rightarrow $ $p^{2}+\xi (b\cdot p)^{2}=0$. We remark
that the nonminimal coupling $B^{\mu }B^{\nu }R_{\mu \nu }$ has not produced
massive modes for the graviton, at least at the linearized level.

It is worthy to note that some constraints here derived are similar to some
gauge fixing conditions arising in the Hilbert-Einstein theory for the
graviton field. Namely, the condition (\ref{eq:constrain1}) is an analogue of
the usual harmonic gauge, $\left( \partial _{\mu }\partial _{\nu }h^{\mu \nu
}-\frac{1}{2}\square h=0\right) $, while the constraint given by Eq. (\ref%
{eq:vin6}) can be interpreted as a kind of axial gauge, as pointed out in
Ref. \cite{BluhmFung}. This situation is similar to the one which happens in
the Proca's model where the Lorenz condition ($\partial _{\mu }A^{\mu }=0$)
appears naturally from the equation of motion, but does not represent a true
gauge fixing condition because Proca's model does not possess a local $U(1)$
gauge invariance. Here, the bumblebee field breaks spontaneously the
diffeomorphism invariance, {i.e., $h_{\mu \nu }\rightarrow h_{\mu \nu
}+\partial _{\mu }\zeta _{\nu }+\partial _{\nu }\zeta _{\mu }$, of the
Hilbert-Einstein gravity. Yet, the graviton remains massless with a modified
dispersion relation.

We should still remark that in limit $\lambda \rightarrow 0$ the potential $V$ vanishes, recovering a gravity theory endowed
with Lorentz and diffeomorphism symmetry, which requires a gauge fixing
condition for the propagator evaluation. This shows that limit $\lambda
\rightarrow 0$ cannot be taken on the Feynman propagator just
carried out. This is similar to the impossibility of taking $m\rightarrow 0
$ in the propagator of a massive graviton to obtain the
propagator of the massless graviton. A route to take such a limit is
the Stuckelberg trick, used to address the van Dam-Veltman-Zakharov
discontinuity in the massless limit of linear massive gravity \cite{Hinter}.

\section{Remarks and conclusions\label{sec:Conclusions}}

In this work, we have investigated the Einstein-Hilbert action in the
presence of the Lorentz-violating bumblebee field. We initially have
considered the linearized version of this theory, including the bumblebee
solution into the interaction term, $\sigma \sqrt{-g}B^{\mu }B^{\nu }R_{\mu
\nu }$. The bilinear bumblebee action in the linearized graviton field $%
h_{\mu \nu }$\ is added to the bilinear linearized Einstein-Hilbert action,
implying the action of our interest. The corresponding Feynman propagator is
exactly carried out using an extended basis of spin projectors based on the
Barnes-Rivers usual ones. The graviton dispersions relations are extracted
from the poles of the propagator and used to study the physical consistency
of the model. It is verified that the graviton possess two
dispersion relations depending explicitly on the bumblebee field vacuum expectation value. The
first one, $p^{2}+\xi (b\cdot p)^{2}=0$, provides massless modes
which are causal (for a positive coupling constant $\xi >0)$ and
unitary (for any values of $b_{0}$ and small values of the
background $\left\vert \mathbf{b}\right\vert )$. The second
dispersion relation, $(b\cdot p)^{2}-b^{2}p^{2}=0$, gives causal
but nonunitary modes which spoil the physical consistency of the model.
Similar nonunitarity issues in Lorentz-violating gravity models were
recently shown in \cite{Pereira2011,hernanski4}.

\begin{acknowledgments}
The authors R. V. Maluf and C. A. S. Almeida are grateful to CAPES and CNPq
for financial support. R. Casana and M. M. Ferreira Jr are grateful to
CAPES, CNPq, and FAPEMA. The authors acknowledge Professor Jos\'{e} A. Helay\"{e}%
l-Neto for useful discussions.
\end{acknowledgments}

\appendix

\section{Barnes-Rivers operators and Lorentz-symmetry breaking \label%
{sec:Appendix-A:-BarnesRiversOperators}}

In the calculation of the inverse kinetic operator, necessary to determine
the graviton propagator, we employ an algorithm based on the Barnes-Rivers
rank-two spin projectors \cite{Barnes,Rivers,Sezgin,Accioly1,Accioly2,Accioly3}%
, given by
\begin{eqnarray}
\text{P}_{\mu\nu,\kappa\lambda}^{(1)} & = &\frac{1}{2}\left( \theta_{\mu
\kappa}\omega_{\nu\lambda}+\theta_{\mu\lambda}\omega_{\nu\kappa}+\theta
_{\nu\kappa}\omega_{\mu\lambda}+\theta_{\nu\lambda}\omega_{\mu\kappa}\right)
,  \notag  \label{eq:P1} \\
\text{P}_{\mu\nu,\kappa\lambda}^{(2)} & = &\frac{1}{2}\left( \theta_{\mu
\kappa}\theta_{\nu\lambda}+\theta_{\mu\lambda}\theta_{\nu\kappa}\right) -%
\frac{1}{3}\theta_{\mu\nu}\theta_{\kappa\lambda},  \notag  \label{eq:P2} \\
\text{P}_{\mu\nu,\kappa\lambda}^{(0-\theta)} & = &\frac{1}{3}\theta_{\mu\nu
}\theta_{\kappa\lambda},\,\,\,\,\,\,\,\,\text{P}_{\mu\nu,\kappa\lambda
}^{(0-\omega)}=\omega_{\mu\nu}\omega_{\kappa\lambda},  \label{eq:Ptheta} \\
\text{{P}}_{\mu\nu,\kappa\lambda}^{(0-\theta\omega)} & = &\frac{1}{\sqrt{3} }%
\left( \theta_{\mu\nu}\omega_{\kappa\lambda}+\theta_{\kappa\lambda}
\omega_{\mu\nu}\right) ,  \notag
\end{eqnarray}
where $\theta_{\mu\nu}=\eta_{\mu\nu}-\omega_{\mu\nu},$\ $\omega_{\mu\nu
}=p_{\mu}p_{\nu}/p^{2}$\ are the transverse and longitudinal projectors,
respectively. The usual spin operators for the subspace of symmetric rank-two
tensors satisfy the following tensorial completeness relation:
\begin{equation}
\left[\text{P}^{(1)}+\text{P}^{(2)}+\text{P}^{(0-\theta)}+\text{P}%
^{(0-\omega)}\right] _{\mu\nu,\kappa\lambda }=\frac{\eta_{\mu\kappa}\eta_{%
\nu\lambda}+\eta_{\mu\lambda}\eta _{\nu\kappa}}{2}.
\end{equation}

The Barnes-Rivers basis was extended to gravity theories involving
Lorentz-symmetry violation in Ref. \cite{Boldo}. In such an extension, the spin
operators induced by the\ Lorentz-violating background $b_{\mu}$ yields
the whole set of structures listed below:
\begin{equation}
{\tilde{\Sigma}}_{\mu\nu}=b_{\mu}p_{\nu}+b_{\nu}p_{\mu},~\ \Lambda_{\mu\nu
}=b_{\mu}b_{\nu},
\end{equation}
\begin{eqnarray}
{\tilde{\Pi}}_{\mu\nu,\kappa\lambda}^{(1)} & = &\frac{\theta_{\mu\kappa} {%
\tilde{\Sigma}}_{\nu\lambda}+\theta_{\mu\lambda}{\tilde{\Sigma}}_{\nu\kappa
}+\theta_{\nu\kappa}{\tilde{\Sigma}}_{\mu\lambda}+\theta_{\nu\lambda} {%
\tilde{\Sigma}}_{\mu\kappa}}{2}, \\
\tilde{\Pi}_{\mu\nu,\kappa\lambda}^{(2)} & = &\frac{\theta_{\mu\kappa}
\Lambda_{\nu\lambda}+\theta_{\mu\lambda}\Lambda_{\nu\kappa}+\theta_{\nu%
\kappa }\Lambda_{\mu\lambda}+\theta_{\nu\lambda}\Lambda_{\mu\kappa}}{2},
\end{eqnarray}
\begin{eqnarray}
\tilde{\Pi}_{\mu\nu,\kappa\lambda}^{(\theta\Sigma)} & =&\frac{\theta_{\mu\nu
}{\tilde{\Sigma}}_{\kappa\lambda}+\theta_{\kappa\lambda}{\tilde{\Sigma}}
_{\mu\nu}}{\sqrt{3}}, \\
\tilde{\Pi}_{\mu\nu,\kappa\lambda}^{(\theta\Lambda)} & = &\frac{%
\theta_{\mu\nu
}\Lambda_{\kappa\lambda}+\theta_{\kappa\lambda}\Lambda_{\mu\nu}}{\sqrt{3}},
\\
\tilde{\Pi}_{\mu\nu,\kappa\lambda}^{(\Lambda\Lambda)}&=&\Lambda
_{\mu\nu}\Lambda_{\kappa\lambda},
\end{eqnarray}
\begin{eqnarray}
\tilde{\Pi}_{\mu\nu,\kappa\lambda}^{(\omega\Lambda-a)} & =
&\omega_{\mu\kappa }{\Lambda}_{\nu\lambda}+\omega_{\mu\lambda}{\Lambda}%
_{\nu\kappa}+\omega _{\nu\kappa}{\Lambda}_{\mu\lambda}+\omega_{\nu\lambda}{%
\Lambda}_{\mu\kappa }, \\
\tilde{\Pi}_{\mu\nu,\kappa\lambda}^{(\omega\Lambda-b)} & = &\omega_{\mu\nu
}\Lambda_{\kappa\lambda}+\omega_{\kappa\lambda}\Lambda_{\mu\nu},
\end{eqnarray}
\begin{eqnarray}
\tilde{\Pi}_{\mu\nu,\kappa\lambda}^{(\omega\Sigma)} & = &\omega_{\mu\nu} {%
\tilde{\Sigma}}_{\kappa\lambda}+\omega_{\kappa\lambda}{\tilde{\Sigma}}
_{\mu\nu}, \\
\tilde{\Pi}_{\mu\nu,\kappa\lambda}^{(\Lambda\Sigma)} & = &\Lambda_{\mu\nu }{%
\tilde{\Sigma}}_{\kappa\lambda}+\Lambda_{\kappa\lambda}{\tilde{\Sigma}}
_{\mu\nu}.
\end{eqnarray}

We should mention that these projectors are not exactly the same ones of
Ref. \cite{Boldo}. Indeed, note that projector $\tilde{\Sigma}_{\mu\nu}$ is
symmetrized while projector $\Sigma_{\mu\nu}$\ of Ref. \cite{Boldo} is
not. The same holds for $\tilde{\Pi}_{\mu\nu,\kappa\lambda}^{(1)},$\ $\tilde{%
\Pi}_{\mu\nu,\kappa\lambda}^{(2)},$\ $\tilde{\Pi}_{\mu\nu
,\kappa\lambda}^{(\omega\Sigma)},$\ $\tilde{\Pi}_{\mu\nu,\kappa\lambda
}^{(\Lambda\Sigma)}.$ Note that the upper indices (1) and (2) in the operators $\tilde{\Pi}_{\mu\nu,\kappa\lambda}^{(1)}$, $\tilde{\Pi}_{\mu\nu
,\kappa\lambda}^{(2)}$ do not refer to its spin content. Some other
interesting ways to defined projectors\ in gravity theories are found in
Refs. \cite{Pereira2011,Sezgin,Hernaski1,Hernaski2,Hernaski3}.


\end{document}